\documentclass{elsart}
\usepackage{graphics}

\usepackage{amssymb}

\begin{document}

\begin{frontmatter}

\title{Production and deceleration of a pulsed beam of metastable NH ($a\,^1\Delta$) radicals}

\author{Sebastiaan Y.T. van de Meerakker,}
\ead{basvdm@fhi-berlin.mpg.de}
\author{Irena Labazan,}
\author{Steven Hoekstra,}
\author{Jochen K\"upper,}
\author{and Gerard Meijer}

\address{Fritz-Haber-Institut der Max-Planck-Gesellschaft\\
         Faradayweg 4-6, 14195 Berlin, Germany}

\begin{abstract}
We report on the production of a pulsed molecular beam of
metastable NH ($a \, ^\textrm{1}\Delta$) radicals and present
first results on the Stark deceleration of the NH ($a \,
^\textrm{1}\Delta, J=\textrm{2}, M\Omega=-\textrm{4}$) radicals
from 550~m/s to 330~m/s. The decelerated molecules are excited on
the spin-forbidden $A\,^\textrm{3} \Pi \leftarrow a\,^\textrm{1}
\Delta$ transition, and detected via their subsequent spontaneous
fluorescence to the $X\,^\textrm{3}\Sigma^{-}, v"=0$ ground-state.
These experiments demonstrate the feasibility of our recently
proposed scheme [Phys. Rev. A 64 (2001) 041401] to accumulate
ground-state NH radicals in a magnetic trap.

\end{abstract}

\end{frontmatter}

\section{Introduction}
\label{Introduction}

In recent years there has been a growing interest in the
development of methods to produce neutral molecules that
are sufficiently slow that they can be confined in a trap.
This interest is triggered by various potential applications
and by the promise of the occurrence of interesting new
physics and chemistry at the low temperatures and high
densities that can ultimately be achieved. Extensive reviews
as well as special issues of scientific journals on the production
and application of cold (polar) molecules have recently appeared
\cite{Bethlem:IntRevPhysChem22:73,specialissue:EPJD31,Krems:IRPC24:99}.

Over the last years, we have developed the method of Stark
deceleration of a molecular beam to produce samples of cold polar
molecules. In a Stark decelerator, the quantum-state specific
force that a polar molecule experiences in an electric field is
exploited to manipulate the external degrees of freedom of a
molecule. This force is rather weak, typically some eight to ten
orders of magnitude weaker than the force that the corresponding
molecular ion would experience in the same electric field. This
force nevertheless suffices to achieve complete control over the
motion of polar molecules, using techniques akin to those used for
the control of charged particles. This has been explicitly
demonstrated by the construction of two types of linear
accelerators \cite{Bethlem:PRL83:1558,Bethlem:PRL88:133003}, a
buncher \cite{Crompvoets:PRL89:093004}, two types of traps
\cite{Bethlem:Nat406:491,Veldhoven:PRL94:083001} and a storage
ring \cite{Crompvoets:Nat411:174} for neutral polar molecules.
Recently, the Stark deceleration and trapping technique has been
used to store ground-state OH radicals in an electrostatic
quadrupole trap for times up to seconds at a density of
$10^7-10^8$ cm$^{-3}$ and at a temperature of 50~mK
\cite{Meerakker:PRL94:23004}.

To be able to study molecular interactions and collective effects
in these trapped samples of polar molecules, the phase-space density
needs to be further increased, i.e., the number density needs to
be made higher and/or the temperature needs to be reduced. In the
Stark deceleration and trap loading method, only a single loading
cycle has been employed thus far. The most straightforward method
to increase the phase-space density of the trapped molecules would
therefore be the accumulation of several packets of decelerated
molecules in the trap. Simply re-loading the trap, however, requires
opening up the trapping potential thereby either losing or heating
the molecules that are already stored. Recently, we proposed a
scheme that circumvents this fundamental obstacle, a scheme that
specifically allows to accumulate ground-state NH radicals in a
magnetic trap \cite{Meerakker:PRA64:041401(R)}.
In this scheme, a beam of NH molecules in the long-lived metastable
$a ^1\Delta$ state is injected into a Stark decelerator and brought
to a standstill. When the molecules are (almost) standing still,
they are optically pumped to the $X \, ^3\Sigma^-$ electronic
ground state by inducing the spin-forbidden
$A\,^3 \Pi \leftarrow a\,^1 \Delta$ transition \cite{Meerakker:PRA68:032508},
followed by spontaneous emission in the triplet system. In the
ground state, the NH molecules can be magnetically trapped, for
instance in a quadrupole magnetic trap. As the Stark interaction
in the ground state is very weak, the electric fields of the
decelerator hardly effect the magnetic trapping potential and
the decelerator can be switched on again to stop and load the
next pulse of NH radicals.

The success and usefulness of the proposed accumulation scheme
depends on the availability of an intense pulsed source of slow
metastable NH ($a \, ^\textrm{1}\Delta$) radicals that is suited
to be coupled to a Stark decelerator. Furthermore, optical pumping
of the decelerated NH ($a \, ^\textrm{1}\Delta$) radicals to their
electronic ground state needs to be performed. In this Letter we
describe a production method for a pulsed molecular beam of
metastable NH ($a \, ^\textrm{1}\Delta$) radicals and in a
preliminary experiment the deceleration of NH radicals from
550~m/s to 330~m/s is demonstrated. The decelerated molecules are
excited to the $A\,^\textrm{3}\Pi$ state and detected via their
subsequent emission to the electronic ground-state. An
experimental procedure that might allow further cooling of a
decelerated packet of NH molecules during the loading of the
magnetic trap is proposed and discussed.

\section{A pulsed slow molecular beam of NH ($\textit{a} \,
^\textrm{1}\Delta$) radicals} \label{sec:NHreloading/slowbeam}

For the deceleration and trapping of NH radicals an intense pulsed
beam of metastable NH radicals with a low initial velocity is
required. Although many cell experiments with NH ($a \,
^1\Delta$), using a variety of production schemes, have been
carried out, only a few studies have been performed with
metastable NH in a pulsed
\cite{Patel-Misra:JCP97:4871,Sauder:JCP91:5316,Gericke:JCP92:6548,Randall:CPL200:113}
or continuous \cite{Ubachs:JMolSpec115:1986,Mo:JCP111:4598}
molecular beam. In most of the beam studies, photo-dissociation of
the precursors HN$_3$, HNCO, or NH$_3$ has been used.

Our experiments are performed in a molecular beam machine that
consists of a differentially pumped source chamber and a detection
chamber that are separated by a 1.5~mm diameter skimmer. The NH
beam is produced by photo-dissociation of hydrazoic acid (HN$_3$)
by focussing the 80~mJ output of a quadrupled Nd:YAG laser at
266~nm just in front of the nozzle orifice of a pulsed supersonic
valve (Jordan Inc.). Using this production scheme, the NH radicals
are almost exclusively produced in the $a \,^1\Delta$ state, since
the NH ($X \, ^3\Sigma^-$) + N$_2$ channel is spin forbidden
\cite{McDonald:CPL51:57,Nelson:JCP93:8777,Hawley:JCP99:2638}. The
HN$_3$ gas is produced online by heating sodium azide in excess of
stearic acid to 95~$^{\circ}$C to drive off the gaseous hydrazoic
acid. In order to reduce the initial velocity of the molecular
beam, either Argon, Krypton or Xenon is co-expanded from the
pulsed valve as inert carrier gas. Unfortunately, Xe is observed
to efficiently quench the NH ($a \,^1\Delta$) radicals, an effect
that has been well documented before
\cite{Patel-Misra:JCP97:4871,Hack:JPC96:47,Adams:JPC95:2975}. The
mechanism that causes the fast quenching rate for Xe is at present
unknown \cite{Mo:JCP111:4598,Dagdigian:AnnuRevPhysChem48:95}. The
quenching rate for Kr is about three orders of magnitude smaller
than for Xe \cite{Hack:jpc:93:3540}, sufficiently low to enable
the use of Kr as a carrier gas in our experiments.

\begin{figure}[!htb]
    \centering
    \resizebox{0.7\linewidth}{!}
    {\includegraphics[0,0][504,400]{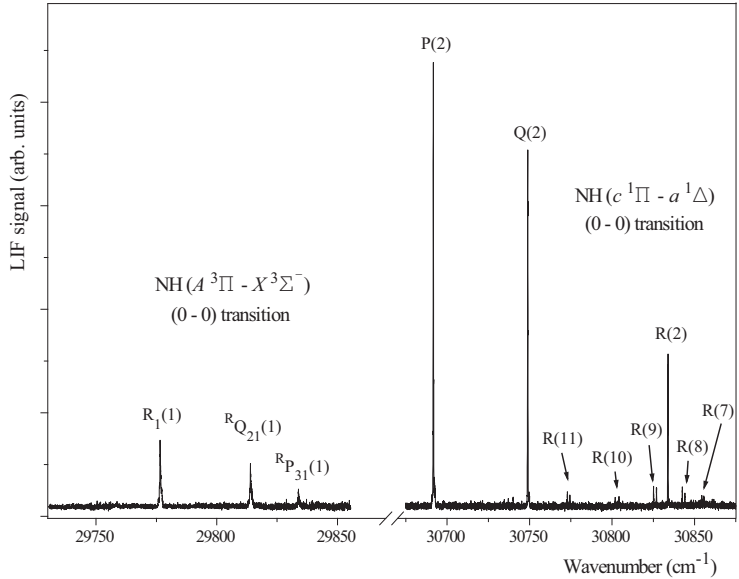}}
    \caption{Fluorescence excitation spectrum of the
    $A \,^3\Pi, v=0 \leftarrow X \,^3\Sigma^-, v=0$ band and the
    $c \,^1\Pi, v=0 \leftarrow a \, ^1\Delta, v=0$ band of NH in the
    pulsed molecular beam. The lines are labelled using standard
    spectroscopic nomenclature.}
    \label{fig:NHreloading/spectrumNHbeam}
\end{figure}

The molecular beam is characterized by detecting the NH radicals
about 24~cm downstream from the nozzle by a resonant Laser Induced
Fluorescence (LIF) detection scheme. The populations in the $a
\,^1\Delta$ and the $X \,^3\Sigma^-$ states are probed by inducing
the $c \,^1\Pi, v=0 \leftarrow a \, ^1\Delta, v=0$ transition
around 326~nm and the $A \,^3\Pi, v=0 \leftarrow X \, ^3\Sigma^-,
v=0$ transition around 336~nm, respectively, and by recording the
molecular fluorescence. The fluorescence excitation spectrum that
is thus obtained is shown in Figure
\ref{fig:NHreloading/spectrumNHbeam}.
The observed spectral lines are labelled using standard
spectroscopic nomenclature. The majority of the NH radicals in the
beam resides in the metastable $a \, ^1\Delta$ electronic state.
The small percentage of the population that is in the electronic
ground state (estimated to be less than 10\%) is believed to
originate from electronic quenching of NH ($a \, ^1\Delta$) by
collisions with Kr, or from two-photon processes in the
dissociation of HN$_3$. In the $a \, ^1\Delta$ state, the
rotational level $J=2$ is the most populated one, reflecting the
efficient rotational cooling in the expansion region of the
supersonic jet. A secondary maximum in the rotational population
distribution is observed around $J=9$. A similar bimodal
distribution has been observed before for NH ($a \, ^1\Delta$) in
a supersonic beam, namely in the production of NH via
photo-dissociation of HN$_3$ at 193~nm in N$_2$ carrier gas
\cite{Sauder:JCP91:5316}. The secondary maximum reflects the
nascent rotational distribution of NH ($a \, ^1\Delta$) upon
photo-dissociation of HN$_3$, and the low inelastic scattering
cross sections for high values of $J$. It is evident from the
spectrum presented in Figure \ref{fig:NHreloading/spectrumNHbeam}
that, in spite of the bimodal rotational level distribution, at
least 50\% of all the NH ($a \, ^1\Delta$, v=0) radicals are in
the J=2 rotational ground-state level.

\begin{figure}[!htb]
    \centering
    \resizebox{0.7\linewidth}{!}
    {\includegraphics[0,0][500,420]{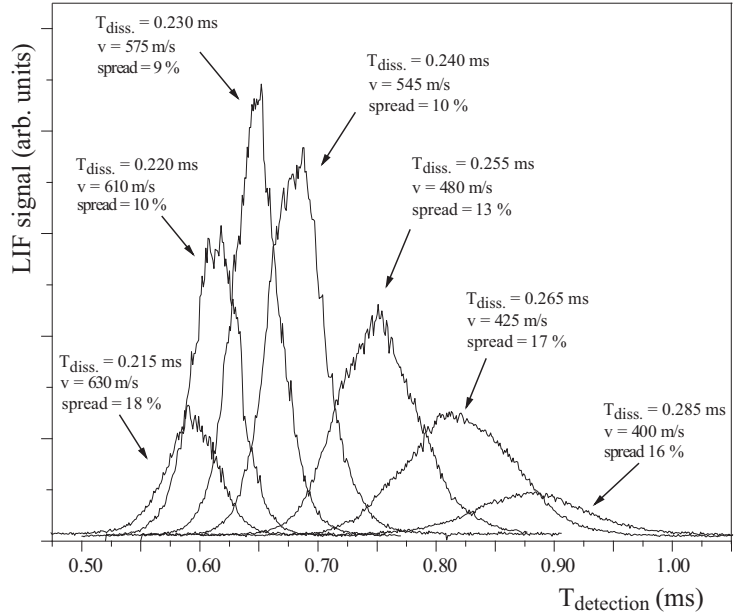}}
    \caption{TOF profiles of the molecular beam of NH
    ($a \, ^1\Delta, J=2$) radicals for different values
    of $T_{diss}$, the time-interval between the trigger for the
    pulsed valve and the firing of the dissociation laser.
    For each TOF profile, the mean velocity and the velocity
    spread (FWHM) of the beam is given.}
    \label{fig:NHreloading/timescansNH}
\end{figure}

The velocity distribution of the molecular beam critically depends
on the time at which the dissociation laser is fired in the gas
pulse that is emanating from the supersonic valve. The
time-interval between the trigger pulse for the pulsed valve and
the trigger pulse for the dissociation laser is indicated by
$T_{diss}$. Here it should be noted that, depending on
experimental conditions, the pulsed valve typically only opens
some 0.20--0.21~ms after the trigger is applied; the dissociation
laser fires simultaneous (on this time-scale) with the opening of
the valve. In Fig. \ref{fig:NHreloading/timescansNH}, the measured
time of flight (TOF) profiles of the beam of NH ($a \, ^1\Delta$)
are shown for different values of $T_{diss}$. The population in
the $J=2$ level is measured by recording the LIF intensity on the
$P(2)$ line of the $c \,^1\Pi \leftarrow a \, ^1\Delta$
transition. For low values of $T_{diss}$, when the valve is just
opening and when there is no good supersonic expansion yet, the
beam typically has a high mean velocity ($>$600~m/s), and a
relatively large velocity spread. The beam has a maximum peak
intensity if the dissociation laser is fired 0.23~ms after the
trigger pulse for the valve. The beam then has the minimum
velocity spread (FWHM) of only 9 \%, at a mean velocity of
575~m/s. The latter is still significantly faster than would be
expected for a room temperature expansion of pure Kr, i.e., $\sim$
450~m/s. This can be explained in part by the seeding of the lower
mass HN$_3$ molecules but mainly reflects the relatively high
constant temperature (70$^{\circ}$C) of the valve body during
operation. In addition, the transient high temperature of the gold
hairpins of the valve opening mechanism, that support a peak
current exceeding 4.0~kA when the valve opens, is known to lead to
even higher apparent source temperatures in this type of valve
\cite{Boogaarts:thesis}. For higher values of $T_{diss}$, the beam
has a lower mean velocity and the inferior expansion conditions
result in a lower peak intensity and a larger velocity spread. The
decreasing beam velocity over the temporal profile of the gas
pulse, is in part due to cooling of the gold hairpins by the gas
that flows in between them and in part due to the transition from
supersonic to effusive flow.

\section{Deceleration of a molecular beam of NH ($a \,^\textrm{1}\Delta$) radicals}
\label{sec:NHreloading/deceleration}

In this section, first experiments are presented in which the
pulsed beam of NH ($a \, ^1\Delta, J=2$) radicals is decelerated.
For this, the Stark decelerator that we employed recently for
the deceleration and trapping of OH radicals
\cite{Meerakker:PRL94:23004} is mounted between the source and
the detection chamber. The general operation principles and the
technical details of this specific Stark decelerator are
documented extensively elsewhere
\cite{Bethlem:PRA65:053416,Meerakker:ARPC:inpress}, and will not
be further detailed here.

After passing through the skimmer, the NH ($a \, ^1\Delta$)
radicals are focussed by a short hexapole into the -- just over
one meter long -- Stark decelerator, consisting out of 102 electric
field stages. In each electric field stage, a voltage difference of
40~kV between opposing electrodes can be applied, creating a maximum
electric field on the molecular beam axis of about 95~kV/cm. The
NH ($a \, ^1\Delta$) radicals that exit the decelerator are
optically pumped to the electronic ground state in the LIF detection
zone, and are state-selectively detected in this process. Excitation
on the $P_2(2)$ line of the spin-forbidden $A \, ^3\Pi, v=0 \leftarrow
a \, ^1\Delta, v=0$ transition around 584~nm \cite{Meerakker:PRA68:032508}
is performed. The $A \, ^3\Pi, v=0 \rightarrow X \, ^3\Sigma^-, v=0$
fluorescence around 336~nm is imaged onto the photomultiplier
(PMT). The detection of the NH radicals in this way is almost
background-free as stray-light from the laser can be effectively blocked
by optical filters in front of the PMT. The transition is
induced by the fundamental output of a ns pulsed dye laser system
(Spectra physics, PDL3 combination). Typically, an energy of 30~mJ
in a 4~mm diameter beam with a bandwidth of 0.04~cm$^{-1}$ is used,
which is not enough to saturate the transition; when a laser with a
superior optical brightness is used, this transition can be readily
saturated \cite{Meerakker:PRA68:032508}.

\begin{figure}[!htb]
    \centering
    \resizebox{0.7\linewidth}{!}
    {\includegraphics[0,0][355,265]{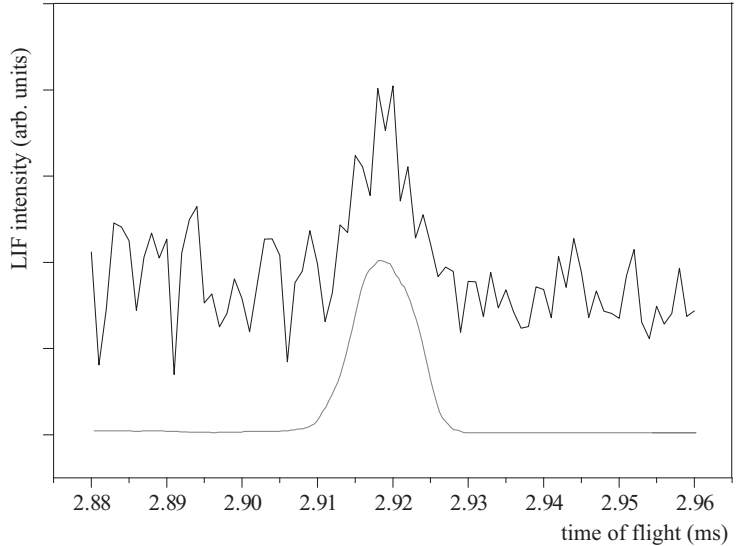}}
    \caption{Observed TOF profile of decelerated NH ($a \, ^1\Delta, J=2, M\Omega=-4$)
    radicals exiting the decelerator (upper curve). The packet of molecules is
    decelerated from 550~m/s to 330~m/s. The TOF profile that results
    from a 3D numerical simulation of the experiment is shown underneath
    the experimental profile.}
    \label{fig:NHreloading/NHdeceleration}
\end{figure}

In Fig. \ref{fig:NHreloading/NHdeceleration} the TOF profile of
the decelerated NH radicals is shown as obtained when the Stark
decelerator is operated at a phase angle $\phi_0 = 70^\circ$ for a
synchronous molecule with an initial velocity of 550~m/s
($E_{kin}=$ 190 cm$^{-1}$). With these settings, the decelerator
extracts about 1.2~cm$^{-1}$ of kinetic energy from the
synchronous molecule in every deceleration stage. NH radicals in
the $M_J\Omega=-4$ component of the $a \, ^1\Delta, J=2$ state are
decelerated to a final velocity of 330~m/s ($E_{kin}=$ 68
cm$^{-1}$), i.e., almost two-thirds of their kinetic energy is
removed. The decelerated bunch of molecules arrives in the
detection region 2.92~ms after production, about 0.6~ms later than
the arrival time of the original non-decelerated molecular beam
(data not shown). The arrival time distribution that results from
a three dimensional trajectory simulation of the experiment is
shown underneath the experimental data. The arrival time of the
packet of decelerated molecules is accurately reproduced by the
simulations. The width of the velocity distribution of the
decelerated NH radicals is about 10 m/s, corresponding to a
longitudinal translational temperature of about 0.1~K.

\section{Cooling of the beam during trap loading}

In the original paper on the accumulation of ground state NH radicals
in a magnetic trap \cite{Meerakker:PRA64:041401(R)}, it was
proposed to transfer the NH radicals with a pulsed laser from
the $a\,^1\Delta$ state to the $X\,^3\Sigma$ state. The strength
of the spin-forbidden $A\,^3\Pi\leftarrow a\,^1\Delta$ transition
\cite{Meerakker:PRA68:032508} suggests, however, that this
transfer can be performed using a cw-laser system as well. In this
case, an increase in the phase-space density of each individual
decelerated packet of NH radicals during the trap loading process
is possible, as outlined below.

\begin{figure}[!htb]
    \centering
    \resizebox{0.7\linewidth}{!}
    {\includegraphics[0,0][507,492]{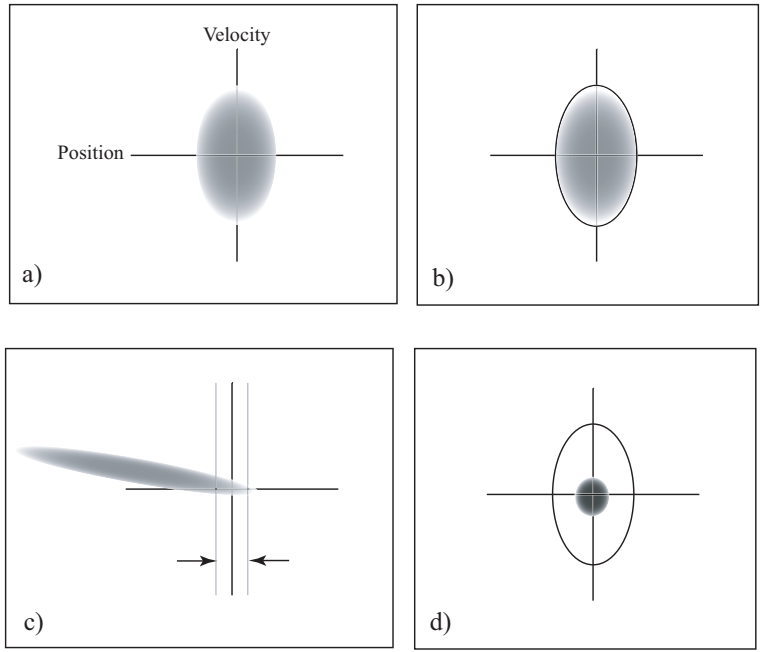}}
    \caption{Idealized longitudinal phase-space distribution at
    the exit of the decelerator (panel a)) with a perfectly
    matched acceptance of a trap (panel b)). A spatially
    elongated, but velocity compressed, longitudinal phase-space
    distribution (panel c)) that can be transferred to the trap
    by a spatially localized laser, resulting in an increased
    phase-space density in the trap (panel d)).}
    \label{fig:NHreloading/phase-space-improvement}
\end{figure}

Whenever a pulsed trap loading scheme is used, either with a
pulsed laser to transfer NH radicals to a magnetic trap or
by abruptly switching on an electrostatic trap, for instance,
the upper limit of the phase-space density that can be achieved in
the trap is set by the phase-space density of the decelerated
packet. This upper limit can be reached by perfectly matching the
(6-dimensional) emittance of the decelerator to the acceptance of
the trap. Imperfect matching, either with respect to the velocity
or spatial acceptance of the trap will lead to a reduced number
of trapped molecules and/or to heating up of the trapped molecules,
e.g., to a reduced phase-space density. The longitudinal phase-space
distribution of the molecular packet at the exit of the decelerator
is, schematically and idealized, shown in panel a) of Figure
\ref{fig:NHreloading/phase-space-improvement}. By abruptly switching
on a trap with a perfectly matched acceptance, this distribution can
be loaded in the trap without loss in phase-space density, as
schematically indicated in panel b).

If one uses a spatially localized, continuous and uni-directional
means to transfer the decelerated molecules to the trap, time
comes in as an extra degree of freedom. The molecules do not have
to arrive at the center of the trap simultaneously in order to be
transferred. This can be used to advantage when, for instance, the
longitudinal spatial distribution of the packet of decelerated molecules
is allowed to expand before entering the trap region, and when
simultaneously the longitudinal velocity spread of the packet is
reduced \cite{Crompvoets:PRL89:093004}. When there is in addition
a small, overall deceleration when the molecules approach the trap
center, the (idealized) longitudinal phase-space distribution as
depicted in panel c) can be realized. In this case, the NH radicals
can be pumped to the magnetic trap by the continuous laser, which
only irradiates near the center of the trap, and their density near
the trap center will thus increase. In panel d) of Figure
\ref{fig:NHreloading/phase-space-improvement} it is shown how the
longitudinal phase-space distribution depicted in panel c) can be
compressed to a higher final phase-space density via this cw-laser
loading scheme.

In order to estimate the increase is phase-space density that can
be accomplished with this scheme we performed preliminary
one-dimensional trajectory calculations of metastable NH radicals
in the decelerator and in the trap-loading region. These simulations
indicate that the width of the final velocity distribution of a packet
of NH radicals transferred with a cw-laser to the magnetic trap is
realistically about a factor of 2.5~smaller than a package of molecules
trapped with the pulsed loading scheme. As indicated in the schematic
description above, this reduction in the width of the velocity distribution
is not the result of simply transferring only the slow molecules; no
molecules are lost during the transfer process. The reduced
velocity spread indicates a potential phase-space density increase
of magnetically trapped NH ($X\,^3\Sigma^-$) radicals of more than
one order of magnitude compared to the trapping of NH ($a\,^1\Delta$)
radicals in a pulsed electrostatic trap, for instance.

\section{Conclusions}

We have demonstrated the production of a pulsed, slow molecular
beam of metastable NH ($a\,^1\Delta$) radicals that is suited to
be injected into a Stark decelerator. In a preliminary
deceleration experiment, metastable NH radicals in the $J=2,
M\Omega=-4$ level have been decelerated from 550~m/s to 330 m/s,
e.g. two-thirds of the kinetic energy has been removed from these
molecules. In principle, the NH radicals can be decelerated to
rest and confined in an electrostatic trap. When the NH ($a
\,^1\Delta$) radicals are transferred to the $X\,^3\Sigma^-$
ground state by inducing the $A\,^3\Pi\leftarrow a\,^1\Delta$
transition when the molecules have come to rest, the ground-state
NH radicals can be confined near the center of a magnetic trap. We
have presented a possible scheme to increase the phase-space
density of the trapped packet of NH based on the use of a cw-laser
for the transfer from the $a\,^1\Delta$ state to the
$X\,^3\Sigma^-$ state. Together with our previously proposed
scheme to accumulate NH radicals from subsequent deceleration
cycles in a magnetic trap, this offers great prospects for further
Stark deceleration and magnetic trapping experiments.

\section{Acknowledgements}

This work is supported by the EU network on "Cold Molecules". The
research of I.L. has been made possible by a Alexander von
Humboldt fellowship.

\end{document}